\documentclass[rnote]{aa}  
\usepackage{longtable,lscape}
\usepackage{graphicx}
\usepackage{lscape}
\usepackage{natbib}
\usepackage{txfonts}
\def\kms{$\rm km\;s^{-1}$}
\def\kmsmpc{$\rm km\;s^{-1}\;Mpc^{-1}$}

\def\ha{H$\alpha$}

\def\niig{[N~{\small II}]$\,\lambda6583$}

\def\siip{[S~{\small II}]$\,\lambda6716$}

\begin{document}
\title{Detection of non-ordered central gas motions in a sample of
four low surface brightness galaxies
\thanks{Based on observations carried out at the European Southern
  Observatory (ESO 71.B-3050)} 
\thanks{Table 2 is only available in electronic form at the CDS via
  anonymous ftp to cdsarc.u-strasbg.fr (130.79.128.5) or via
  http://cdsweb.u-strasbg.fr/Abstract.html} } 
\titlerunning{Non-ordered gas motion in the center of LSB galaxies.}
\authorrunning{Pizzella et al.}

\author{A. Pizzella\inst{1}
\and D. Tamburro\inst{2}
\and E.M. Corsini\inst{1, 3}
\and F. Bertola\inst{1} }

\offprints{alessandro.pizzella@unipd.it}

\institute{Dipartimento di Astronomia, Universit\`a di Padova, 
  vicolo dell'Osservatorio~3, I-35122 Padova, Italy 
\and Max-Planck-Institut f\"ur Astronomie, K\"onigstuhl 17, 
  D-69117 Heidelberg, Germany 
\and Scuola Galileiana di Studi Superiori, via VIII Febbraio 2, 
  35122 Padova, Italy}

\date{Accepted January 14, 2008}

\abstract 
{}
{We present integral-field spectroscopy of the ionized gas in the
central regions of four galaxies with a low surface brightness disk
taken with the Visible Multi Object Spectrograph at the Very Large
Telescope and aimed at testing the accuracy in the determination of
the central logarithmic slope $\alpha$ of the mass density radial
profile $\rho(r)\propto r^\alpha$ in this class of objects.}
{For all the sample galaxies we subtracted from the observed velocity
field the best-fit model of gas in circular motions and derived the
residuals. Only ESO-LV~5340200 is characterized by a regular velocity
field. We extracted the velocity curves of this galaxy along several
position angles, in order to estimate the uncertainty in deriving the
central gradient of the total mass density from long-slit
spectroscopy.}
{We report the detection of strong non-ordered motions of the ionized
gas in three out of four sample galaxies. The deviations have velocity
amplitudes and spatial scales that make not possible to disentangle
between cuspy and core density radial profiles.}
{}


\maketitle

\section{Introduction}
\label{sec:introduction}

Low surface brightness (LSB) galaxies are defined as disk galaxies
with a central face-on brightness fainter than 22.6 $B-$mag
arcsec$^{-2}$. In comparison with high surface brightness (HSB)
galaxies, the LSB galaxies have higher mass-to-light ratios and are
dominated by the dark matter (DM) even in the central regions
\citep[e.g.,][]{Debl1996,Cour1999,Borr2001}. Therefore, the rotation 
curves of LSB galaxies are thought to represent an ideal test-bed to
check the predictions of $N-$body simulations in a cold dark matter
(CDM) universe, where the mass density radial profile of the DM halos
is described by a steep power law $\rho(r)\sim r^\alpha$
\citep[i.e. $\alpha\lesssim-1$,][]{Nava1997,Nava2004,Moore1999,Diem2005}.

On the other hand, the observations seem to be inconsistent with the
CDM scenario. The DM mass density distribution derived by H~{\small
I}\ rotation curves of most of LSB galaxies is described by a radial
profile with a constant core profile
\citep[i.e. $\alpha\simeq0$,][]{mcgaugh1998,salu2001}. This is true
also for the ionized-gas rotation curves
\citep[see][]{Mcga2001,Debl2001Let,Mcga2003}. The spatial resolution
of these data allows to detect cuspy DM density profiles when they are
present. But, other authors do find an agreement of the radio
\citep{vdbosch2000} and optical \citep{Swat2003} data with the CDM
predictions. Despite this observational effort, a unique answer to the
central DM density profile slope has not been found.

However, measuring the mass density profile using the gas as potential
tracer in galactic centers may result problematic not only because of
the resolution effects. Indeed, the gas can have an intrinsic velocity
dispersion and does not necessarily move on perfectly ballistic orbits
\citep{Bert1995,Cinz1999}. Moreover, the emissivity distribution of the gas 
is often clumpy \citep{Vand2001} and the presence of asymmetric and/or
decoupled structures could remain undetected in long-slit observations
\citep[see][and references therein]{Cocc2005}. More mundanely, the imperfect 
centering and orientation of the slit in optical spectroscopy and the
beam smearing in radio observations contribute to the total
uncertainties on the measured kinematics.

To address all these issues, it is crucial to obtain high-resolution
gas kinematics along different axes. Only recently, the
two-dimensional velocity field of the ionized gas in LSB galaxies has
been measured by means of integral-field units spectroscopy.
\citet{Simo2005} studied a sample of low mass spiral galaxies 
with a velocity field which can be explained by either a cored or a
cuspy density radial profile. \citet{Kuzi2006} observed a sample of
LSB galaxies. They found that DM halos with a core of constant mass
density provide a better fit to the data with respect to those with a
density cusp.

In this paper we focus on the intrinsic limitation of the ionized-gas
kinematics in tracing the galactic central potential and mass
distribution. The work is based on the kinematics in the center of
four galaxies with a LSB disk measured by integral-field spectroscopy.
The observations and data reduction are described in
Sect. \ref{sec:observations}. The measurement and analysis of the
kinematics and distribution of the ionized gas are discussed in
Sect. \ref{sec:analysis}. The results are given in
Sect. \ref{sec:results}. Finally, the conclusions are presented in
Sect. \ref{sec:conclusions}.

\begin{figure*}
\includegraphics[width=.42\textwidth]{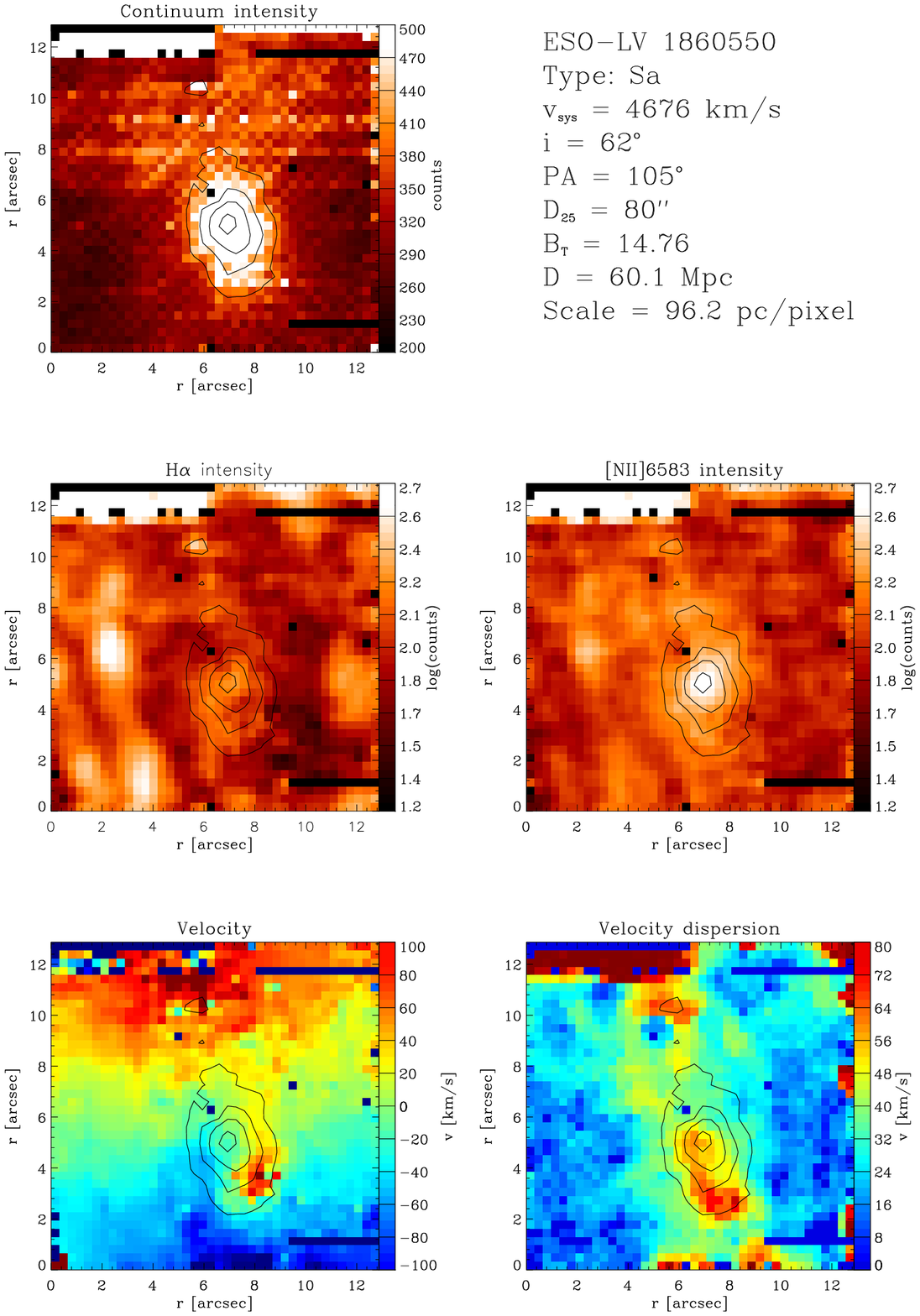}
\includegraphics[width=.42\textwidth]{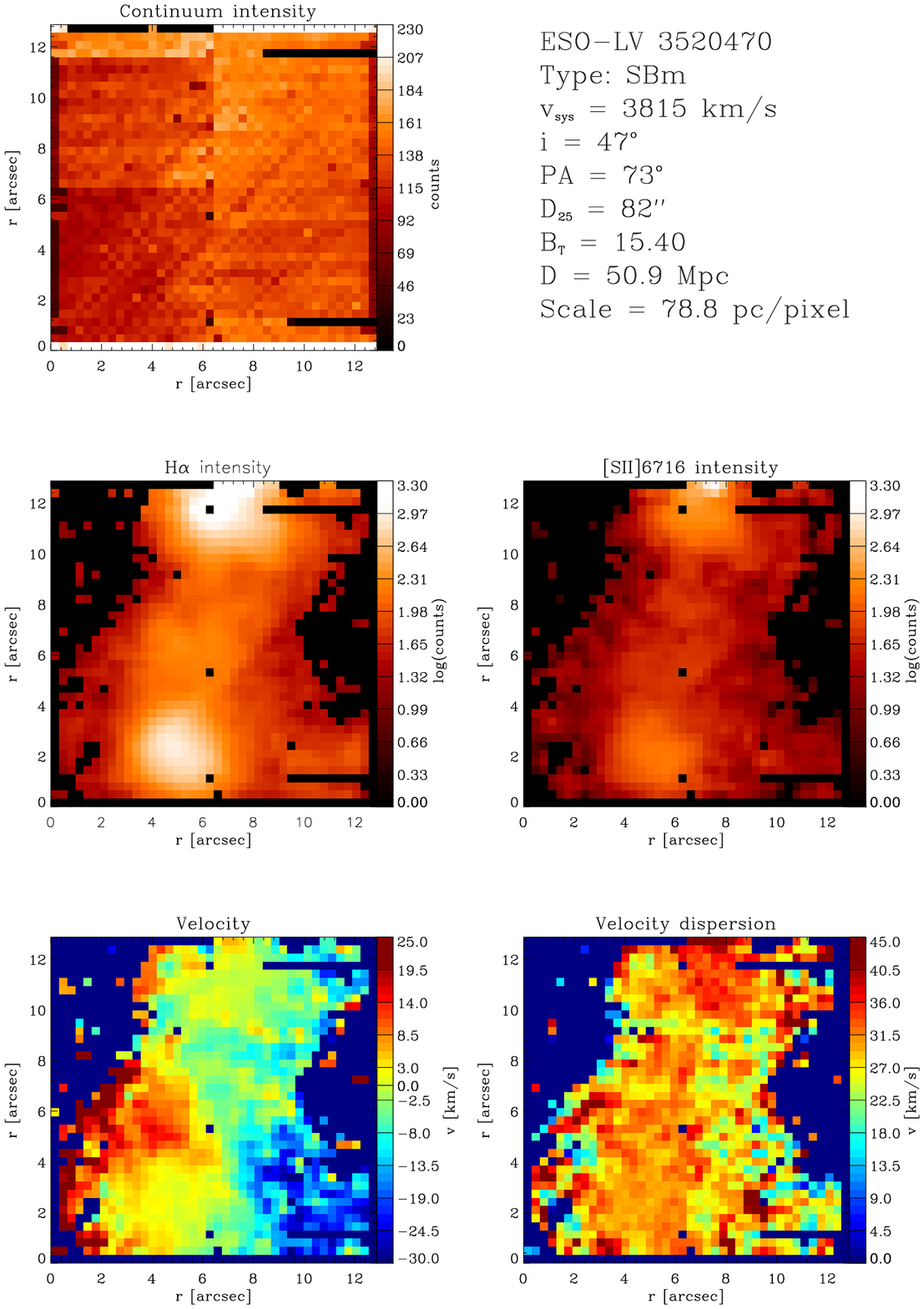}
\includegraphics[width=.42\textwidth]{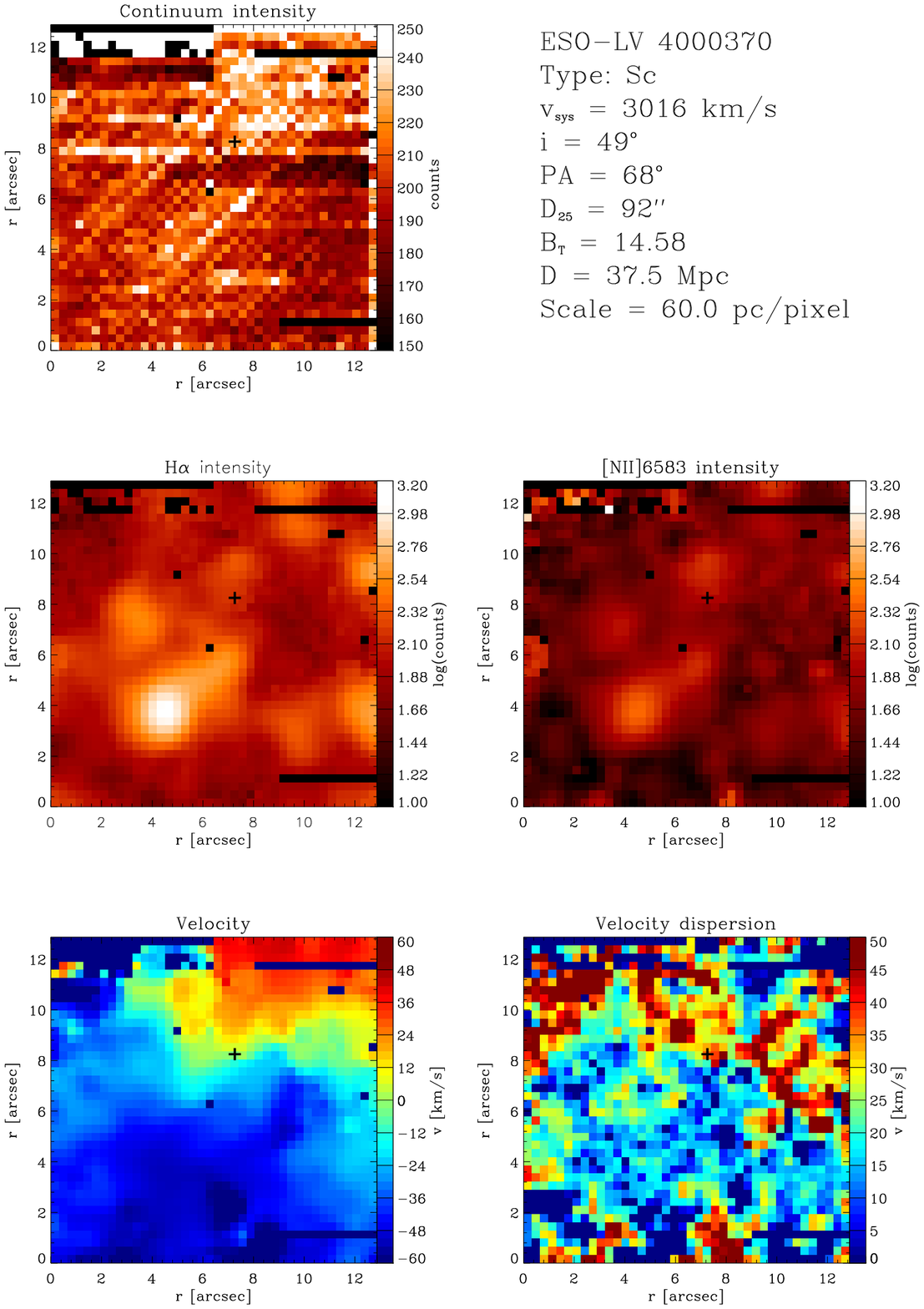}
\hfill
\includegraphics[width=.42\textwidth]{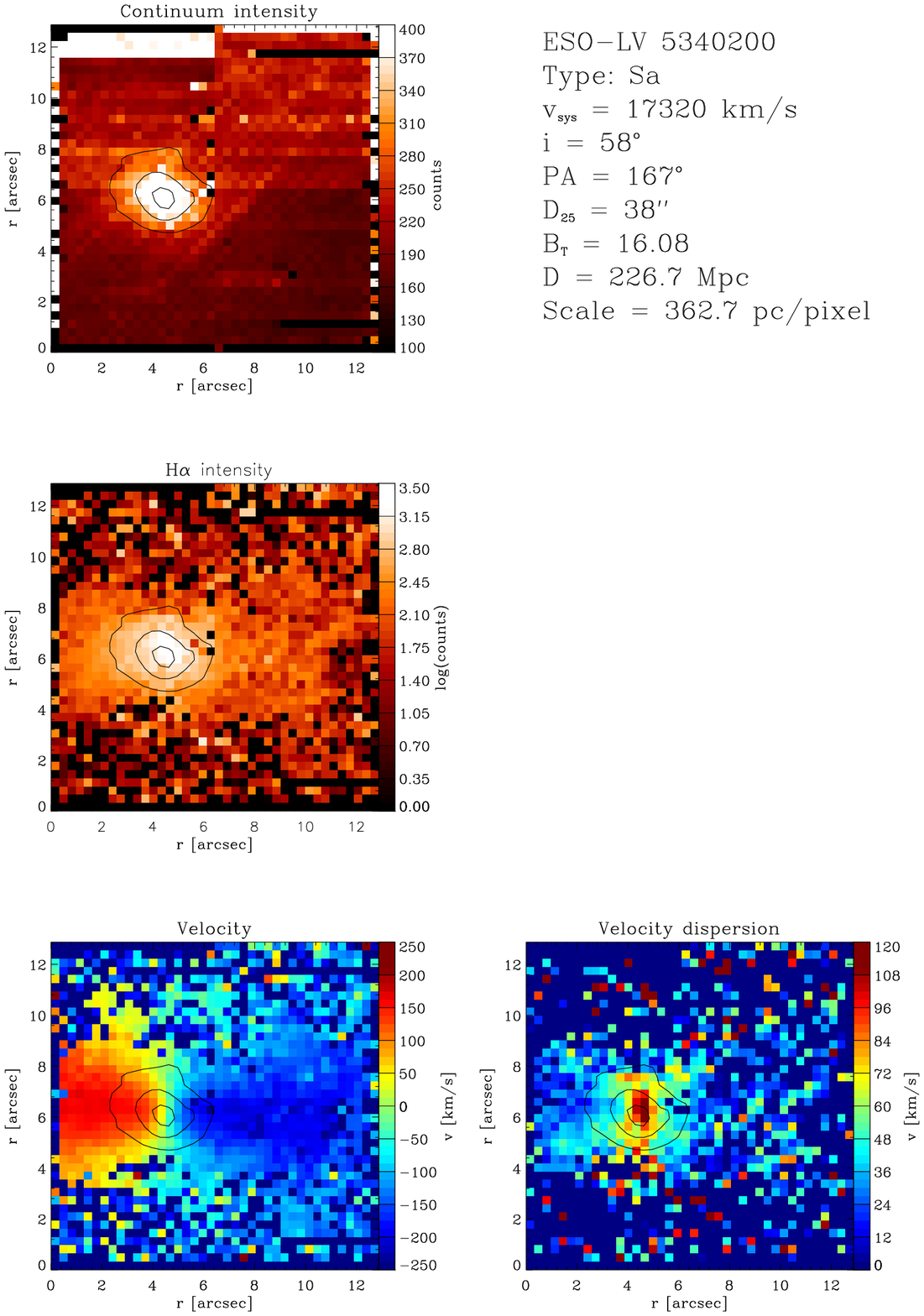}

\caption{Maps of the distribution and kinematics of the ionized gas
for the sample galaxies. East is up and North is right. The ranges are
indicated at the right of each panel. {\em Upper panel:} reconstructed
image of the galaxy obtained by integrating the stellar continuum
between 5728 and 5758 \AA . {\em Middle left panel:} \ha\ intensity
map. {\em Middle rigth panel:} \niig\ (or \siip ) intensity map. {\em
Lower left panel:} heliocentric line-of-sight velocity without
applying any correction for galaxy inclination. {\em Lower right
panel:} line-of-sight velocity dispersion corrected for instrumental
velocity dispersion. The morphological classification, inclination,
major-axis position angle, diameter of the 25 $B-$mag arcsec$^{-2}$
isophote, and total blue magnitude are taken from
\citet[][ESO-LV]{ESOLV}. The systemic velocities are from this paper.
The distances are derived from the systemic velocities corrected to
the CMB reference frame following
\citet{Fixsen1996} and assuming $H_0=75$ \kmsmpc .}
\label{fig:results}
\end{figure*}

\section{Observations and data reduction} 
\label{sec:observations}

The sample galaxies were selected among those observed by
\citet{Mcga2001} and \citet{Pizz2005..LSB} to have a LSB disk and
measured ionized-gas kinematics with long-slit spectroscopy. Their
main properties are reported in Fig.~\ref{fig:results}. 

The integral-field spectroscopic observations were carried out with
the Very Large Telescope (VLT) at the European Southern Observatory
(ESO) in Paranal (Chile) in 2003. The Unit Telescope 3 mounted the
Visible Multi Object Spectrograph (VIMOS) in the Integral Field Unit
(IFU) configuration.  The field of view of the four VIMOS channels
(Q1-Q4) was $13''\times 13''$ and it was projected onto a microlenses
array. This was coupled to optical fibers which were rearranged on a
linear set of microlenses to produce an entrance pseudoslit to the
spectrograph.  The pseudoslit generated a total of 1600 spectra
covering the field of view with a spatial resolution of $0\farcs33$
per fiber. Each channel was equipped with the HR\_orange high
resolution ($R \sim 2500$) grism and a thinned and back illuminated
EEV44 CCD with $2048\times4096$ pixels of $15\times15$ $\mu$m$^2$. The
wavelength range between 5250 and 7450 \AA\ was covered with a
reciprocal dispersion is 0.65 \AA\ pixel$^{-1}$.
The observing log and values of the full width at half maximum
of the seeing as measured by the ESO Differential Image Meteo Monitor
are reported in Table~\ref{tab:log}.

\begin{table}[h!]
\caption{Observing log}
\begin{tabular}{llcl}
\hline
\noalign{\smallskip}
\multicolumn{1}{c}{Name} & \multicolumn{1}{c}{Date} & 
\multicolumn{1}{c}{Exp. Time} & \multicolumn{1}{c}{FWHM}  \\
\noalign{\smallskip}
\hline             
\noalign{\smallskip}
\object{ESO-LV 1860550}&  07/08 Apr 2003 & $2\times45$ min &$0\farcs4/0\farcs6$  \\
\object{ESO-LV 3520470}&  31 Jul 2003    & $1\times45$ min &N/A      \\
\object{ESO-LV 4000370}&  24 Aug 2003    & $1\times45$ min &$0\farcs5$      \\
\object{ESO-LV 5340200}&  28/29 Jun 2003 & $2\times45$ min &$0\farcs7/0\farcs6$  \\
\noalign{\smallskip}
\hline
\end{tabular}
\label{tab:log} 
\end{table}

For each VIMOS channel all the spectra were traced, identified, bias
subtracted, flatfield corrected, corrected for relative fiber
transmission, and wavelength calibrated using the routines of the
VIPGI pipeline \citep{Scodeggio2005}. The cosmic rays and bad pixels
were identified and cleaned using our routines developed under the IDL
environment\footnote{The Interactive Data Language is a product of
Research Systems, Inc. (RSI)}.
We checked and corrected the wavelength rebinning by measuring the
difference between the measured and predicted wavelength for the 23
brightest night-sky emission lines in the observed spectral range
\citep{Oste1996}. The rms of the differences was $0.08$ \AA\ (Q1),
$0.05$ \AA\ (Q2), $0.05$ \AA\ (Q3), and $0.07$ \AA\ (Q4),
corresponding to an accuracy in wavelength calibration of $3$ \kms\ in
the observed spectral range.
The intensity of the night-sky emission lines was used to correct for
the different relative transmission of the VIMOS channels.
The width of the night-sky emission lines was used to estimate the
instrumental line width. We measured $\rm FWHM=1.86\pm0.19$ \AA\ (Q1),
$1.81\pm0.19$ \AA\ (Q2), $1.88\pm0.14$ \AA\ (Q3), and $1.84\pm0.14$
\AA\ (Q4), corresponding to a instrumental velocity dispersion
$\sigma_{\rm inst} = 36$ \kms.

\section{Analysis}
\label{sec:analysis}

The ionized-gas kinematics was measured by fitting the brightest
emission lines in the galaxy spectra. They were the \ha\ and \niig\
lines for ESO-LV~1860550 and ESO-LV~4000370, \ha\ and \siip\ lines for
ESO-LV~3520470, and \ha\ line for ESO-LV~5340200. A Gaussian profile
and a straight line were fitted to each emission line and its adjacent
continuum, respectively. The lines were assumed to share a common
centroid velocity and a common velocity width. The heliocentric
correction was applied to the fitted line-of-sight velocities. They
were not corrected for the galaxy inclination. The fitted
line-of-sight velocity dispersions were corrected for the instrumental
velocity dispersion. No flux calibration was performed.
The resulting kinematics of the ionized gas as well as the intensity
maps of the stellar continuum and fitted emission lines are plotted in
Fig. \ref{fig:results}. The resulting line-of-sight heliocentric
velocities and velocity dispersions are listed in Tab.~2.

\begin{figure}
\vspace{0pt}
\includegraphics[width=.24\textwidth]{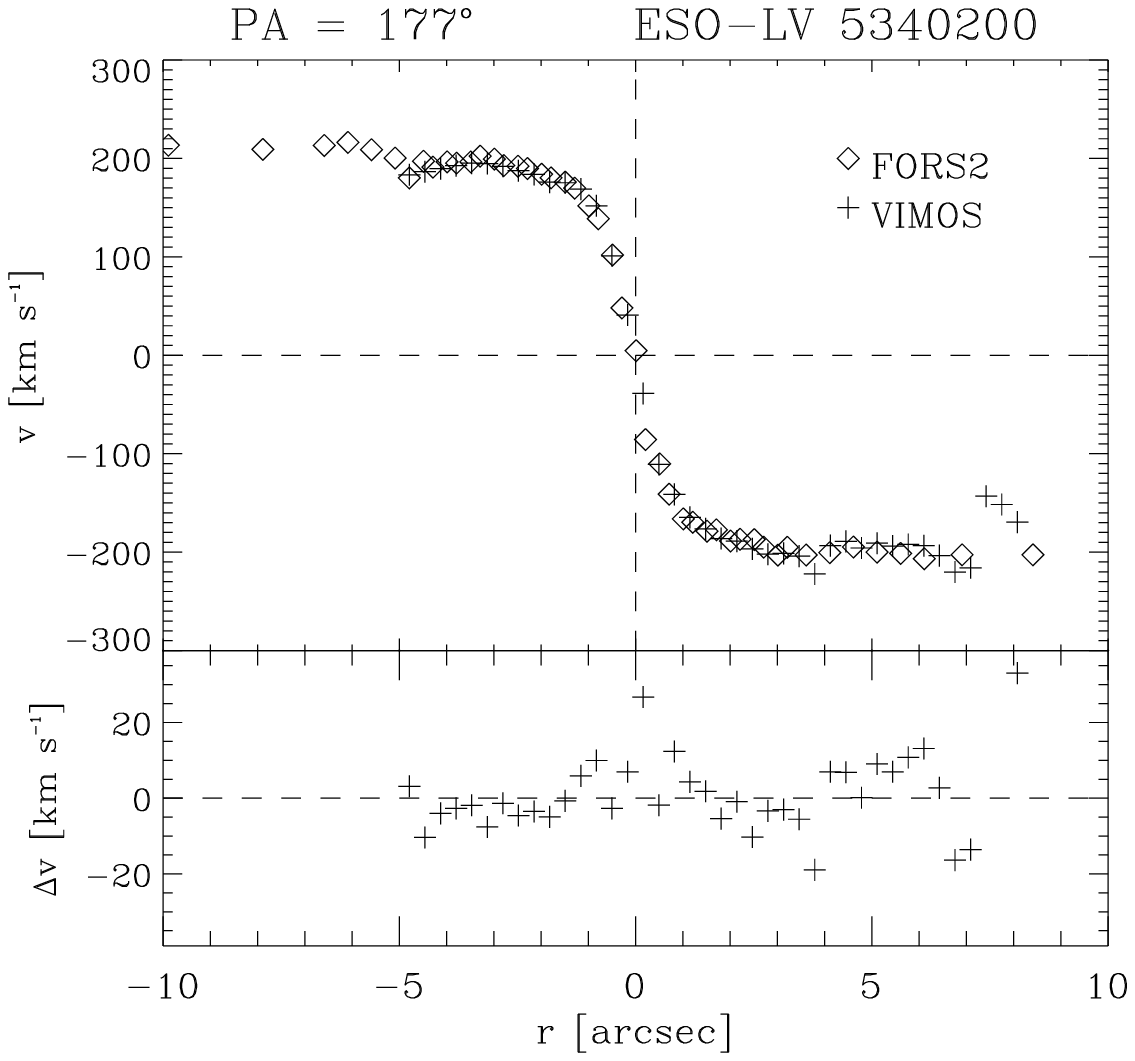}
\hfill
\includegraphics[width=.24\textwidth]{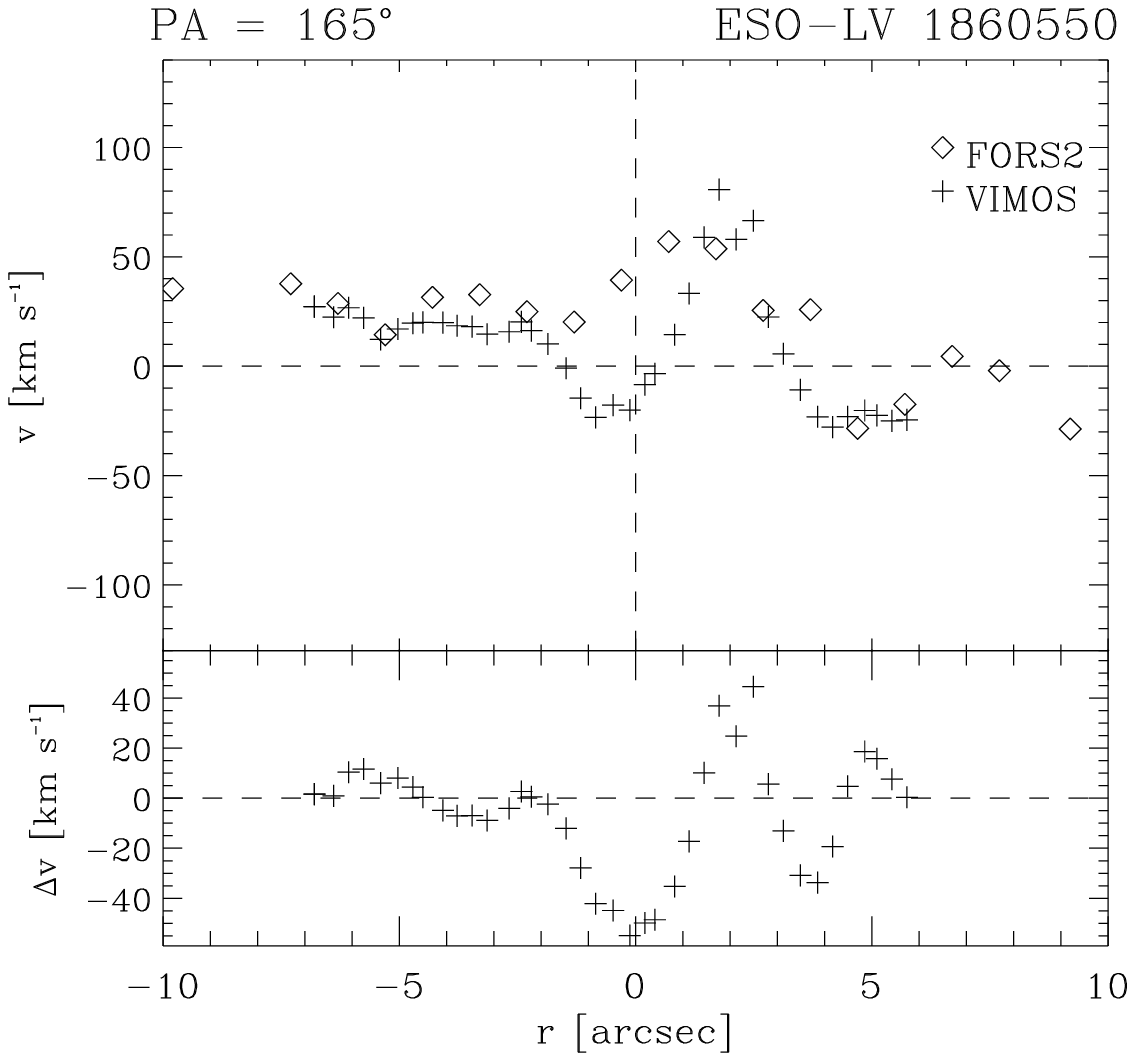}
\caption{The rotation curves measured in this paper ({\em  crosses})
along the major-axis of ESO-LV~5340200 ({\em left panel}) and a
diagonal axis of ESO-LV~1860550 ({\em right panel}) compared with
those obtained from long-slit spectra by \citet[][{\em
diamonds}]{Pizz2005..LSB}.}
\label{fig:comparison}
\end{figure}

\citet{Mcga2001} measured the gas kinematics along the major axis of
ESO-LV~3520470. \citet{Pizz2005..LSB} measured the gas kinematics
along several axes of the remaining three galaxies. A comparison with
these datasets was performed to assess the accuracy and reliability of
our measurements.
We extracted from our two-dimensional velocity fields the velocity
curves along the same axes observed by \citet{Mcga2001} and
\citet{Pizz2005..LSB} mimicking their instrumental setup. 
For all the galaxies we found an agreement with the errors between the
velocity curves extracted from the two-dimensional velocity fields and
the long-slit measurements. The best match was found for the
major-axis rotation curve of ESO-LV~5340200
(Fig. \ref{fig:comparison}).  The largest deviations ($\Delta v
\simeq40$ \kms ) were observed along a diagonal axis ($\rm
PA=165^\circ$) of ESO-LV~1860550 (Fig. \ref{fig:comparison}) and they
can be attributed to the different point spread functions and spatial
samplings of the two datasets.

For each galaxy, we fitted the observed velocity field with the model
of a thin disk of rotating gas to investigate the presence of
non-circular and non-ordered motions. They limit the accuracy of the
mass density distribution derived from the available kinematics.
The model of the gas velocity field is generated assuming that the
ionized-gas component is moving onto circular orbits in an
infinitesimally thin disk with a negligible velocity dispersion.
We assume that the circular velocity $v_{\rm disk}$ at a given radius
$r$ of the gaseous disk is
\begin{equation} 
v_{\rm disk}(r) = v_{\infty}\,\left[ 1-\frac{h}{r} 
  \arctan \frac{r}{h} \right]^{\frac{1}{2}} ,
\label{eq:vcirc}
\end{equation}
where $v_{\infty}$ and $h$ are the asymptotic velocity and a scale
radius, respectively. Following \citet{Cocc2007}, the ionized-gas
velocity measured along the line of sight at a given sky point $(x,y)$
is
\begin{equation}
v_{\rm los}(x,y) = v_{\rm disk}(r) \sin{i} \cos{\phi} + v_{\rm sys} ,
\label{eq:vlos}
\end{equation}
where $v_{\rm sys}$ is the systemic velocity of the galaxy. The
anomaly $\phi$ is measured on the disk plane and it is defined by 
\begin{equation} 
\cos\phi=\left[ (x-x_{0}) \cos \theta + (y-y_{0}) \sin \theta \right] / R ,
\label{eq:anomaly}
\end{equation}
where $(x_{0},y_{0})$, $i$, and $\theta$ are the coordinates of the
center, inclination, and position angle of the line of nodes of the
gaseous disk, respectively. 
The parameters of our model are the asymptotic velocity, velocity
scale radius, and position angle of the gaseous disk, and the systemic
velocity of the galaxy. The disk inclination was not fitted because
the field of view was too small to properly constrain it. We adopted
the inclination given by the ESO-LV catalog
(Fig. \ref{fig:results}). The center of the gaseous disk was assumed
to be coincident with the position of the intensity peak of the
stellar continuum of the reconstructed image for ESO-LV~1860550 and
ESO-LV~5340200 (Fig. \ref{fig:results}). No intensity peak was
observed for ESO-LV~3520470 and ESO-LV~4000370. Therefore, for these
galaxies $x_{0}$ and $y_{0}$ were left as free parameters.
The best-fitting parameters were obtained by iteratively fitting a
model velocity field to the observed one using a non-linear
least-squares minimization method. It is based on the robust
Levenberg-Marquardt method implemented by \citet{More1980}. The actual
computation was done using the MPFIT algorithm under the IDL
environment. The seeing effects were taken into account by convolving
the model with a Gaussian kernel with a FWHM as given in
Tab.~\ref{tab:log}. Finally, the model was compared to the observed
velocity field excluding bad or dead fibers.
We were not able to construct any reliable disk model for
ESO-LV~3520470, because its velocity field turned out to be very
irregular.

Assuming a spherical mass distribution, the mass density profile is 
given by   
\begin{equation}
\rho(r)=\frac{1}{4\pi G}\; \left[2\,\frac{v_{\rm c}}{r}\,\frac{dv_{\rm c}}{dr} + 
\left( \frac{v_{\rm c}}{r} \right)^{2} \right] ,
\label{eq:density} 
\end{equation} 
where $v_{\rm c}$ is the circular velocity
\citep[e.g.,][]{Debl2001Let,Swat2003}.
To avoid possible biases, we derived $v_{\rm c}$ by deprojecting the
line-of-sight velocities $v_{\rm los}$ measured along different axes
without adopting any parametric function (e.g., Eq. \ref{eq:vcirc}) to
describe the velocity field. Finally, we fitted the resulting mass
density profiles with a power law $\rho(r)\sim r^{\alpha}$. The
logarithmic slope $\alpha$ was derived excluding the data points
inside the seeing disk (Tab.~\ref{tab:log}).
ESO-LV~5340200 was the only sample galaxy with a velocity field
suitable for such a kind of analysis. The gas velocity fields of the
other three galaxies were dominated by non-circular and non-ordered
motions.

\begin{figure*}[ht!]
\vspace{0pt}
\includegraphics[width=.33\textwidth]{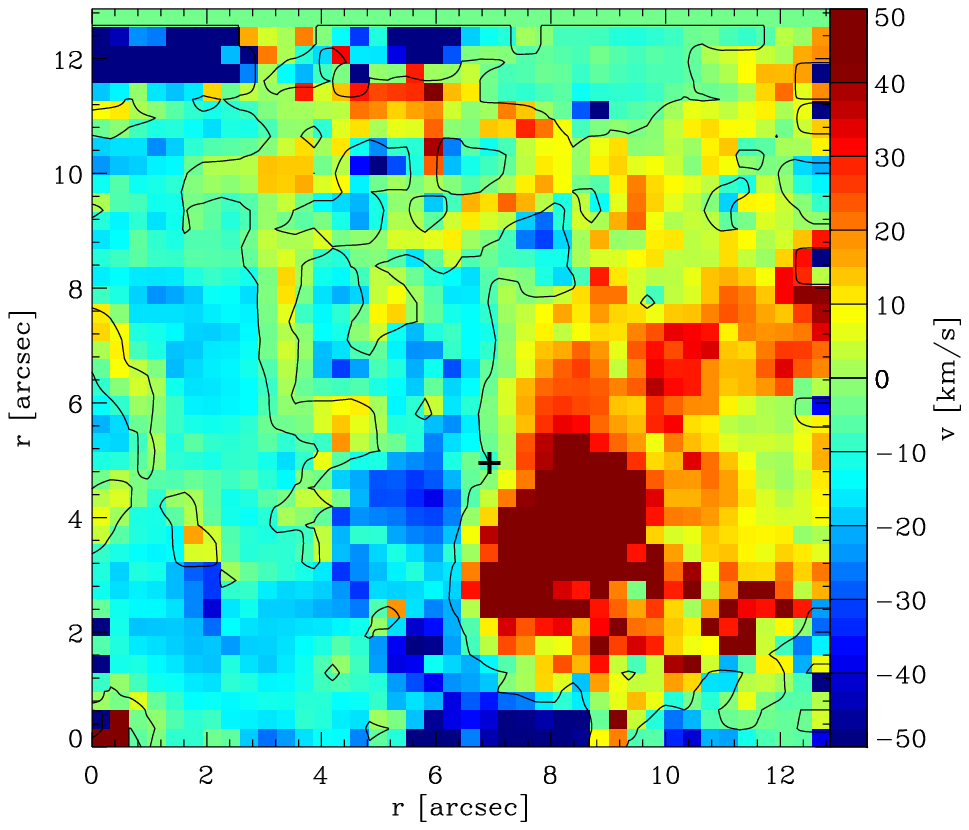}
\hfill
\includegraphics[width=.33\textwidth]{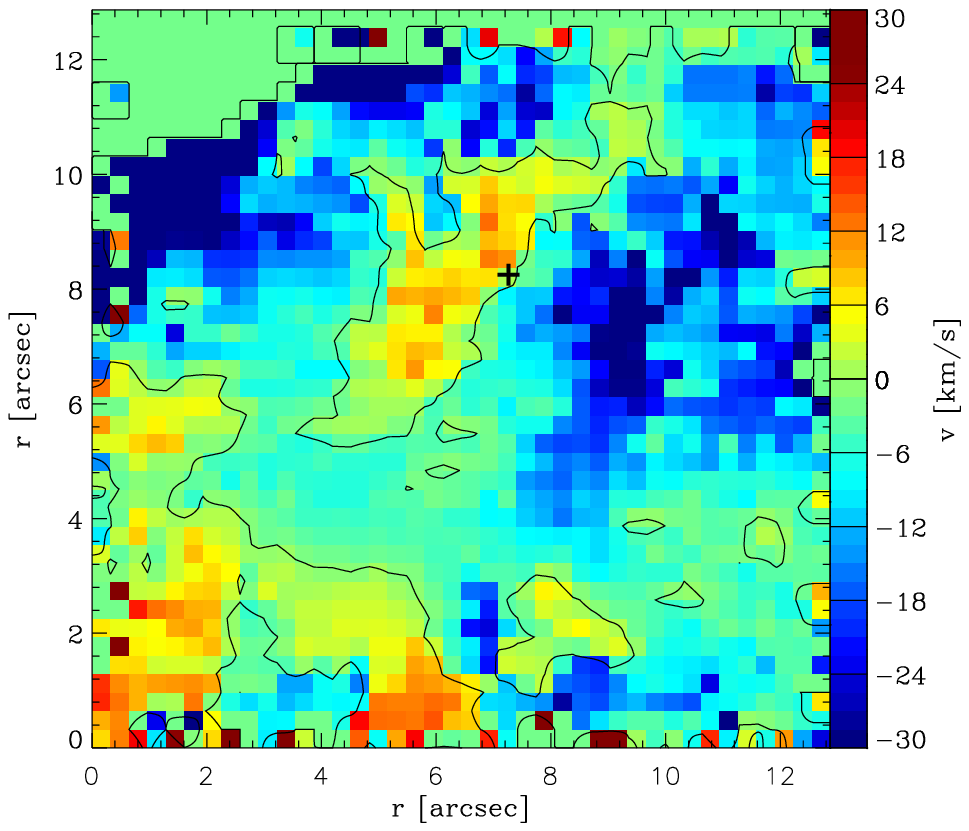}
\hfill
\includegraphics[width=.33\textwidth]{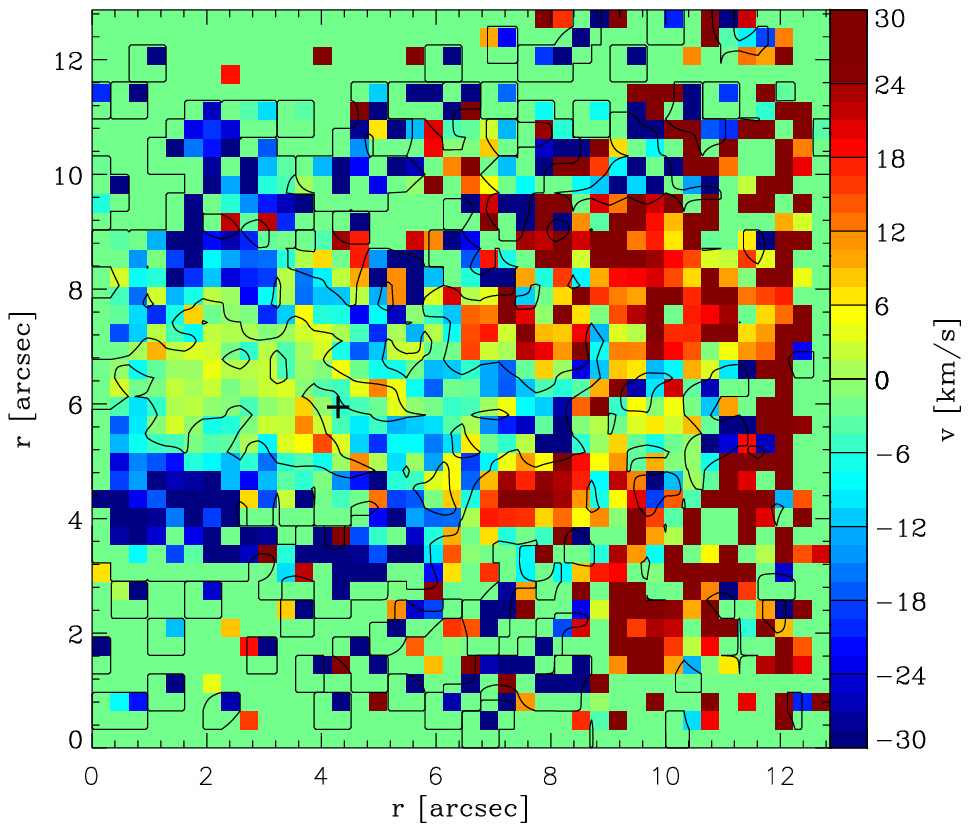}
\caption{Map of the velocity residuals of ESO-LV~1860550
({\em left panel}), ESO-LV~4000370 ({\em central panel}), and
ESO-LV~5340200 ({\em right panel}) after subtracting the best-fit
circular velocity field. A cross indicates the kinematic center.}
\label{fig:residuals}
\end{figure*}

\section{Results}
\label{sec:results}

\subsection{ESO-LV~1860550}
\label{sec:result186}

The field of view covers $3.8 \times 3.8$ kpc$^2$ at the galaxy
distance. The intensity and kinematic maps derived from the two
exposures available for this galaxy are in agreement. 
The intensity maps of the \ha\ and \niig\ lines are shown in
Fig.~\ref{fig:results}.  They unveil the presence in the center of the
field of view of a region with a size of about $7''\times 14''$ and
oriented along the galaxy major axis, where $I($\niig$)/I($\ha$)\geq2$. 
At larger radii the ratio drops to 0.7 and the distribution of the
ionized gas becomes more patchy with the emission regions aligned in a
sort a spiral arm structure. We argued that these features are due to
gas exhaustion by massive stellar formation and to shocks induced by
radiation and winds, as found in the center of NGC~7331 by
\citet{battaner03}.

The velocity field exhibits the overall shape of a rotating disk
(Fig.~\ref{fig:results}). However, there are signatures of non-ordered
motions in the inner few arcsec. They are clearly visible in the
residual map (Fig.~\ref{fig:residuals}) obtained by subtracting the
circular velocity model from the observed velocity field. In
particular, there is a structure receding with $v\simeq120$ \kms ,
which is characterized by the highest velocity dispersion measured in
the frame ($\sigma \simeq 80$ \kms). It is located in the southeastern
quadrant of the field of view at about $3''$ from the galactic
center. In the same region, we measured a narrow ($\sigma \simeq 50$
\kms ) \ha\ absorption line. It is likely that it is due to a
(relatively cold) hydrogen cloud, which is located between the
observer and the galaxy and falling towards the galactic center.  This
region was masked in calculating the circular velocity model to
minimize systematic errors. Except for such a remarkable feature, the
velocity dispersion field is regular and symmetric. The velocity
dispersion peaks at about $60$ \kms\ in very center and then it
decreases outwards.
The asymmetry and irregularity of the observed velocity field, which
are possibly due to an ongoing acquisition event, prevented us to
derive the radial profile of the mass density.

\subsection{ESO-LV~3520470}
\label{sec:result352}

The field of view covers $3.2 \times 3.2$ kpc$^2$ at the galaxy
distance. We excluded the \niig\ line in measuring the kinematics and
distribution of the ionized-gas in ESO-LV~3520470, since it was in
blend with a night-sky emission line.
The \ha\ and \siip\ intensity maps show that the emitting regions are
aligned along the galaxy major axis (Fig.~\ref{fig:results}). The
ionized-gas distribution is characterized by three prominent blobs: a
fainter one in the galactic center and two brighter ones at the two
sides.

There is no velocity gradient along the galaxy major axis. This is
consistent with the long-slit observations in the same wavelength
range by \citet{Mcga2001}. They found that the gas rotation is very
low ($v\leq10$ \kms) out to $20''$. We measured a $\Delta v \simeq 50$
\kms\ over a radial range of $16''$ (corresponding to about 4 kpc)
along a direction close to the galaxy minor axis ($\rm
PA=160^\circ$). The misaligned velocity gradient is not associated to
any feature of the intensity maps, since its center of symmetry is off
set by about $2''$ westward with respect to the central intensity
peak. These features could be interpreted as due to the presence of a
kinematically-decoupled component, similar to the inner polar disks
found in a number of early-type disk galaxies
\citep{Cors2003,Cocc2005,Sil2006}. 
After masking the central region, negligible rotation velocities are
observed. Therefore, we were not able to find the kinematic center and
the radial profile of the mass density was not derived.

\subsection{ESO-LV~4000370}
\label{sec:result400}

The field of view covers $2.4 \times 2.4$ kpc$^2$ at the galaxy
distance. The center of the field does not correspond to the kinematic
center, which is located in the northeastern quadrant. It is marked
with a cross in Fig.~\ref{fig:results}.
The \ha\ and \siip\ intensity maps show a clumpy distribution of the
ionized gas. An intensity peak was observed in the southwestern
quadrant at about $6''$ from the galactic center, which is
characterized by a low emission intensity.

The velocity field, although it is not entirely visible, is not
characterized by the expected pattern for a regular rotating disk. In
fact, it displays an S-shaped distortion in the nucleus and rotation
around the galaxy minor axis. The central kinematically-decoupled
component is seen in the residual map (Fig.~\ref{fig:residuals})
obtained by subtracting the circular velocity model from the observed
velocity field, after masking the central regions. It has a size of
about $4''$ (corresponding to 0.7 kpc) and a $\Delta v \simeq 60$ \kms
.  Its position angle is $\rm PA=150^\circ$, whereas the position
angle of the galaxy major axis is $\rm PA=68^\circ$
The velocity dispersion reaches in the center its maximum value of
about $45$ \kms , and it is mildly correlated with the
$I($\niig$)/I($\ha$)$ ratio. Its value is about 1 in the very center
and it falls to less than 0.5 at larger radii.
The small number of data points, due to the poor centering of the
galaxy on the field of view, did not allow to derive the central
mass density distribution for this object.

\subsection{ESO-LV~5340200}
\label{sec:result534}

This is the farthest galaxy of the sample. It was included in our list
due to the wrong systemic velocity ($cz=3551$ \kms) reported in the
ESO-LV catalog. We measured $v_{\rm sys} = 17320$ \kms\ and derived a
distance of 227 Mpc. Therefore, the field of view covers $14 \times
14$ kpc$^2$.
The two exposures available for this galaxy differ in the telescope
pointing. The intensity and kinematic maps derived from them are in
agreement. This ensures that the data reduction was successfully
performed.  We excluded the \niig\ line in measuring the kinematics
and distribution of the ionized-gas in ESO-LV~5340200, since it was in
blend with two night-sky emission lines.
The distribution of the ionized gas seen in the \ha\ intensity map is
similar to that of the stars showed in the reconstructed image of the
stellar continuum (Fig.~\ref{fig:results}).

The velocity field of the ionized gas exhibits the overall shape of a
rotating disk (Fig.~\ref{fig:results}). ESO-LV~5340200 has the most
regular field of the sample galaxies. The rms is 8 \kms\ in the
central $2\times3$ kpc$^2$ of the residual image obtained by
subtracting the circular velocity model from the observed velocity
field (Fig.~\ref{fig:residuals}). Only kinematic sub-structures with a
size smaller than 0.5 kpc remain undetected. For a comparison, we
measured the rms in the residual map of ESO-LV~1860550 after rebinning
it to the spatial sampling of ESO-LV~5340200. We found a rms of 25
\kms\ (26 \kms\ before the rebinning).
The velocity dispersion is about $40$ \kms . Only in the central
fibers it peaks the a maximum value of about $120$ \kms . This is due
to the combined effect of the limited spatial resolution and sharp
velocity gradient.

\begin{figure}
\vspace{0pt}
\includegraphics[width=9.0cm]{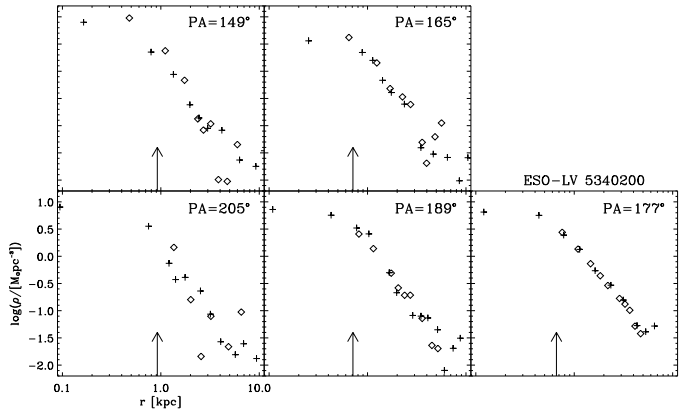}
\caption{The mass density profiles along the major
($\rm PA=177^\circ$) and four diagonal axes of ESO-LV~5340200. In each
panel the diamonds and crosses correspond to data measured along the
approaching and receding side of the galaxy, respectively. The arrow
marks the size of the seeing disk.}
\label{fig:density}
\end{figure}

Since the ionized gas of ESO-LV~5340200 is characterized by ordered
motions, we used its velocity field as a test case to study the
typical uncertainties in estimating the mass density profile by means
of long-slit spectroscopy.
We divided the disk in five $22.5^\circ-$wide wedges. The first wedge
is oriented along the line of nodes, two wedges are oriented along
directions which are $\pm 22.5^\circ$ offset from the line of nodes,
and the last two wedges are oriented along directions which are $\pm
45^\circ$ offset from the line of nodes. According to galaxy
inclination and orientation, on the sky plane the wedges are oriented
along directions which are $\pm 12.4^\circ$ and $\pm 27.9^\circ$ from
the galaxy kinematic major axis ($\rm PA=177^\circ$).
We extracted the velocity curve in each wedge of the two-dimensional
velocity field. Ten velocity curves were derived, since the
approaching and receding sides of the galaxy were independently
considered. For each curve, we computed the term $dv_{\rm c}/dr$ on
adjacent points. The density profiles were hence derived from
Eq.~\ref{eq:density} and plotted in Fig.~\ref{fig:density}. 
The logarithmic slope $\alpha=1.2\pm0.3$ was determined at a distance
from the center equal to the seeing disk size, where we were confident
that the data were not affected by beam smearing effects. The size of
the deprojected seeing disk increases for directions drifting away
from the major axis as shown in Fig.~\ref{fig:density}. 
The scatter in the measurements is due to random errors in the
$\alpha$ determination. We did not find any trend with the position
angle as we would expect if the non-circular motions induced by a
triaxial component were present.
It is worth noticing that the galaxy distance and the presence of a
bright bulge \citep{Pizz2005..LSB} do not allow to derive the mass
density distribution of the DM component without applying a full
dynamical model \citep[e.g.,]{Cors1999,Pign2001}. The velocity field
of ESO-LV~5340200 is different from those of LSB galaxies analyzed in
measuring the central content and distribution of DM to address the
core/cusp problem \citep[e.g.,][]{Debl2001Let,Mcga2003,Kuzi2006}.  In
particular, the central velocity gradients of ESO-LV~5340200 was
derived with larger uncertainties, and therefore the value of $0.3$
can be considered an upper limit to the typical uncertainty in the
estimation of logarithmic slope $\alpha$.

\section{Discussion and conclusions}
\label{sec:conclusions}

The results of the study of the sample galaxies can be summarized into
two remarkable aspects, concerning the presence of non-ordered motions
of the ionized gas and the uncertainties of recovering the logarithmic
slope of the radial profile of the mass density in galaxies with a
regular kinematics.

In three objects we found the evidence for both non-ordered motions
and kinematically-decoupled components. ESO-LV~3520470 and
ESO-LV~4000370 host a small ($\sim1$ kpc) structure which is rotating
around the galaxy minor axis. In ESO-LV~1860550 we found the evidence
for a gaseous component infalling toward the galactic center.
There is no strong correlation between the distribution and kinematics
of the ionized-gas of these galaxies and their starlight
distribution. Indeed, there is no hint for the presence of the
decoupled components in broad-band images
\citep[see][]{Pizz2005..LSB}.

These kinematic disturbances seriously affect the determination of the
radial profile of the mass density. In fact, the typical sizes of
these features (0.5--1 kpc) are comparable to the radial range where
the differences between the core and cuspy density radial profiles are
expected to be observed. Their velocity amplitudes (30--50 \kms) are
also of the same order of both the observed rotation velocities and
velocity differences inferred by the different mass density profiles.
The majority of previous studies on the central mass distribution of
LSB galaxies were based on long-slit spectroscopy, where the presence
of kinematic irregularities could have been undetected. 
Non-circular motions are common phenomenon in inner regions of disk
galaxies. Indeed, \citet{Cocc2004} found that about $50\%$ of the
unbarred bright galaxies show strong off-plane and non-circular gas
motions in their centers.

ESO-LV~5340200 was the only sample galaxy characterized by a regular
velocity field. It was too distant to recover the mass density radial
profile in the central kpc with the desired accuracy. However, it
represented a good test-bed to estimate the uncertainties of the
measurement of the central velocity gradient. This is crucial in
deriving the logarithmic slope $\alpha$ of the mass density radial
profile in objects displaying regular kinematics. We found
$\alpha=1.2\pm0.3$. The error was obtained by fitting a power law to
the density distribution derived from the different rotation curves
extracted along several position angles.  Therefore, it represents a
good estimate of the scatter obtained by deriving $\alpha$ using the
gas kinematic measured from long-slit spectroscopy. On the other hand,
the central velocity gradient of ESO-LV~5340200 is larger than that of
typical LSB galaxies. We concluded the error we derived can be
considered an upper limit to the typical uncertainty in the estimation
of logarithmic slope of the mass density.
Although it may be substantially reduced with the analysis of the full
two-dimensional velocity field \citep{Simo2005,Kuzi2006}, the presence
of non-circular and non-ordered gas motion remains an unresolved
issue. A way to circumvent it is using the stellar kinematics. The
measurement of the stellar line-of-sight velocity distribution 
\citep{Pizz2005..LSB} allows to build reliable mass models providing a 
valid alternative to the usual gas-based mass distribution
determinations \citep{Pizz2008, Magorrian2008}.

\begin{acknowledgements} 
The data published in this paper were reduced using the VIMOS
Interactive Pipeline and Graphical Interface (VIPGI) designed by the
VIRMOS Consortium. We thank Bianca Garilli for her support in using
VIPGI and hospitality at the INAF-IASF Milano. This work was made
possible through grants PRIN 2005/32 by Istituto Nazionale di
Astrofisica (INAF) and CPDA068415/06 by Padua University.
\end{acknowledgements}


\begin{thebibliography}{35}
\expandafter\ifx\csname natexlab\endcsname\relax\def\natexlab#1{#1}\fi

\bibitem[{{Battaner} {et~al.}(2003){Battaner}, {Mediavilla}, {Guijarro},
  {Arribas}, \& {Florido}}]{battaner03}
{Battaner}, E., {Mediavilla}, E., {Guijarro}, A., {Arribas}, S., \& {Florido},
  E. 2003, \aap, 401, 67

\bibitem[{{Bertola} {et~al.}(1995){Bertola}, {Cinzano}, {Corsini}, {Rix}, \&
  {Zeilinger}}]{Bert1995}
{Bertola}, F., {Cinzano}, P., {Corsini}, E.~M., {Rix}, H., \& {Zeilinger},
  W.~W. 1995, \apjl, 448, L13

\bibitem[{{Borriello} \& {Salucci}(2001)}]{Borr2001}
{Borriello}, A. \& {Salucci}, P. 2001, \mnras, 323, 285

\bibitem[{{Cinzano} {et~al.}(1999){Cinzano}, {Rix}, {Sarzi}, {Corsini},
  {Zeilinger}, \& {Bertola}}]{Cinz1999}
{Cinzano}, P., {Rix}, H.-W., {Sarzi}, M., {et~al.} 1999, \mnras, 307, 433

\bibitem[{{Coccato} {et~al.}(2005){Coccato}, {Corsini}, {Pizzella}, \&
  {Bertola}}]{Cocc2005}
{Coccato}, L., {Corsini}, E.~M., {Pizzella}, A., \& {Bertola}, F. 2005, \aap,
  440, 107

\bibitem[{{Coccato} {et~al.}(2007){Coccato}, {Corsini}, {Pizzella}, \&
  {Bertola}}]{Cocc2007}
{Coccato}, L., {Corsini}, E.~M., {Pizzella}, A., \& {Bertola}, F. 2007, \aap,
  465, 777

\bibitem[{{Coccato} {et~al.}(2004){Coccato}, {Corsini}, {Pizzella}, {Morelli},
  {J.~G.~Funes S.}, \& {Bertola}}]{Cocc2004}
{Coccato}, L., {Corsini}, E.~M., {Pizzella}, A., {et~al.} 2004, \aap, 416, 507

\bibitem[{{Corsini} {et~al.}(2003){Corsini}, {Pizzella}, {Coccato}, \&
  {Bertola}}]{Cors2003}
{Corsini}, E.~M., {Pizzella}, A., {Coccato}, L., \& {Bertola}, F. 2003, \aap,
  408, 873

\bibitem[{{Corsini} {et~al.}(1999){Corsini}, {Pizzella}, {Sarzi}, {Cinzano},
  {Vega Beltr{\' a}n}, {Funes}, {Bertola}, {Persic}, \& {Salucci}}]{Cors1999}
{Corsini}, E.~M., {Pizzella}, A., {Sarzi}, M., {et~al.} 1999, \aap, 342, 671

\bibitem[{{Courteau} \& {Rix}(1999)}]{Cour1999}
{Courteau}, S. \& {Rix}, H. 1999, \apj, 513, 561

\bibitem[{{de Blok} {et~al.}(2001){de Blok}, {McGaugh}, {Bosma}, \&
  {Rubin}}]{Debl2001Let}
{de Blok}, W.~J.~G., {McGaugh}, S.~S., {Bosma}, A., \& {Rubin}, V.~C. 2001,
  \apjl, 552, L23

\bibitem[{{de Blok} {et~al.}(1996){de Blok}, {McGaugh}, \& {van der
  Hulst}}]{Debl1996}
{de Blok}, W.~J.~G., {McGaugh}, S.~S., \& {van der Hulst}, J.~M. 1996, \mnras,
  283, 18

\bibitem[{{Diemand} {et~al.}(2005){Diemand}, {Zemp}, {Moore}, {Stadel}, \&
  {Carollo}}]{Diem2005}
{Diemand}, J., {Zemp}, M., {Moore}, B., {Stadel}, J., \& {Carollo}, C.~M. 2005,
  \mnras, 364, 665

\bibitem[{{Fixsen} {et~al.}(1996){Fixsen}, {Cheng}, {Cottingham}, {Folz},
  {Inman}, {Kowitt}, {Meyer}, {Page}, {Puchalla}, {Ruhl}, \&
  {Silverberg}}]{Fixsen1996}
{Fixsen}, D.~J., {Cheng}, E.~S., {Cottingham}, D.~A., {et~al.} 1996, \apj, 470,
  63

\bibitem[{{Kuzio de Naray} {et~al.}(2006){Kuzio de Naray}, {McGaugh}, {de
  Blok}, \& {Bosma}}]{Kuzi2006}
{Kuzio de Naray}, R., {McGaugh}, S.~S., {de Blok}, W.~J.~G., \& {Bosma}, A.
  2006, \apjs, 165, 461

\bibitem[{{Lauberts} \& {Valentijn}(1989)}]{ESOLV}
{Lauberts}, A. \& {Valentijn}, E.~A. 1989, {The Surface Photometry Catalogue of
  the ESO-Uppsala Galaxies} (European Southern Observatory, Garching)

\bibitem[{{Magorrian} {et~al.}(2008){Magorrian}, {Sarzi}, {Corsini},
  {Pizzella}, {M\'endez Abreu}, \& {Bertola}}]{Magorrian2008}
{Magorrian}, J., {Sarzi}, M., {Corsini}, E.~M., {et~al.} 2008, \mnras , in
  preparation

\bibitem[{{McGaugh} {et~al.}(2003){McGaugh}, {Barker}, \& {de Blok}}]{Mcga2003}
{McGaugh}, S.~S., {Barker}, M.~K., \& {de Blok}, W.~J.~G. 2003, \apj, 584, 566

\bibitem[{{McGaugh} \& {de Blok}(1998)}]{mcgaugh1998}
{McGaugh}, S.~S. \& {de Blok}, W.~J.~G. 1998, \apj, 499, 41

\bibitem[{{McGaugh} {et~al.}(2001){McGaugh}, {Rubin}, \& {de Blok}}]{Mcga2001}
{McGaugh}, S.~S., {Rubin}, V.~C., \& {de Blok}, W.~J.~G. 2001, \aj, 122, 2381

\bibitem[{{Moore} {et~al.}(1999){Moore}, {Quinn}, {Governato}, {Stadel}, \&
  {Lake}}]{Moore1999}
{Moore}, B., {Quinn}, T., {Governato}, F., {Stadel}, J., \& {Lake}, G. 1999,
  \mnras, 310, 1147

\bibitem[{{Mor{\'e}} {et~al.}(1980){Mor{\'e}}, {Garbow}, \&
  {Hillstrom}}]{More1980}
{Mor{\'e}}, J.~J., {Garbow}, B.~S., \& {Hillstrom}, K.~E. 1980, {User guide for
  MINPACK-1} (Argonne National Laboratory Report ANL-80-74)

\bibitem[{{Navarro} {et~al.}(1997){Navarro}, {Frenk}, \& {White}}]{Nava1997}
{Navarro}, J.~F., {Frenk}, C.~S., \& {White}, S.~D.~M. 1997, \apj, 490, 493

\bibitem[{{Navarro} {et~al.}(2004){Navarro}, {Hayashi}, {Power}, {Jenkins},
  {Frenk}, {White}, {Springel}, {Stadel}, \& {Quinn}}]{Nava2004}
{Navarro}, J.~F., {Hayashi}, E., {Power}, C., {et~al.} 2004, \mnras, 349, 1039

\bibitem[{{Osterbrock} {et~al.}(1996){Osterbrock}, {Fulbright}, {Martel},
  {Keane}, {Trager}, \& {Basri}}]{Oste1996}
{Osterbrock}, D.~E., {Fulbright}, J.~P., {Martel}, A.~R., {et~al.} 1996, \pasp,
  108, 277

\bibitem[{{Pignatelli} {et~al.}(2001){Pignatelli}, {Corsini}, {Vega
  Beltr{\'a}n}, {Scarlata}, {Pizzella}, {Funes}, {Zeilinger}, {Beckman}, \&
  {Bertola}}]{Pign2001}
{Pignatelli}, E., {Corsini}, E.~M., {Vega Beltr{\'a}n}, J.~C., {et~al.} 2001,
  \mnras, 323, 188

\bibitem[{{Pizzella} {et~al.}(2008{\natexlab{a}}){Pizzella}, {Corsini},
  {Sarzi}, {Magorrian}, {Mend\'ez Abreu}, {Coccato}, {Morelli}, \&
  {Bertola}}]{Pizz2005..LSB}
{Pizzella}, A., {Corsini}, E., {Sarzi}, M., {et~al.} 2008{\natexlab{a}}, \mnras
  , submitted

\bibitem[{{Pizzella} {et~al.}(2008{\natexlab{b}}){Pizzella}, {Corsini},
  {Sarzi}, {Magorrian}, \& {Bertola}}]{Pizz2008}
{Pizzella}, A., {Corsini}, E.~M., {Sarzi}, M., {Magorrian}, J., \& {Bertola},
  F. 2008{\natexlab{b}}, in Formation and Evolution of Galaxies, ed. J.~G.
  {Funes, S.J.} \& E.~M. {Corsini} (ASP, San Francisco), in press

\bibitem[{{Salucci}(2001)}]{salu2001}
{Salucci}, P. 2001, \mnras, 320, L1

\bibitem[{{Scodeggio} {et~al.}(2005){Scodeggio}, {Franzetti}, {Garilli},
  {Zanichelli}, {Paltani}, {Maccagni}, {Bottini}, {Le Brun}, {Contini},
  {Scaramella}, {Adami}, {Bardelli}, {Zucca}, {Tresse}, {Ilbert}, {Foucaud},
  {Iovino}, {Merighi}, {Zamorani}, {Gavignaud}, {Rizzo}, {McCracken}, {Le
  F{\`e}vre}, {Picat}, {Vettolani}, {Arnaboldi}, {Arnouts}, {Bolzonella},
  {Cappi}, {Charlot}, {Ciliegi}, {Guzzo}, {Marano}, {Marinoni}, {Mathez},
  {Mazure}, {Meneux}, {Pell{\`o}}, {Pollo}, {Pozzetti}, \&
  {Radovich}}]{Scodeggio2005}
{Scodeggio}, M., {Franzetti}, P., {Garilli}, B., {et~al.} 2005, \pasp, 117,
  1284

\bibitem[{{Sil'Chenko}(2006)}]{Sil2006}
{Sil'Chenko}, O.~K. 2006, in ASSL Vol. 337: Astrophysical Disks, ed. A.~V.
  {Fridman}, M.~Y. {Marov}, \& I.~G. {Kovalenko} (Springer, Dordrecht), 275

\bibitem[{{Simon} {et~al.}(2005){Simon}, {Bolatto}, {Leroy}, {Blitz}, \&
  {Gates}}]{Simo2005}
{Simon}, J.~D., {Bolatto}, A.~D., {Leroy}, A., {Blitz}, L., \& {Gates}, E.~L.
  2005, \apj, 621, 757

\bibitem[{{Swaters} {et~al.}(2003){Swaters}, {Verheijen}, {Bershady}, \&
  {Andersen}}]{Swat2003}
{Swaters}, R.~A., {Verheijen}, M.~A.~W., {Bershady}, M.~A., \& {Andersen},
  D.~R. 2003, \apjl, 587, L19

\bibitem[{{van den Bosch} {et~al.}(2000){van den Bosch}, {Robertson},
  {Dalcanton}, \& {de Blok}}]{vdbosch2000}
{van den Bosch}, F.~C., {Robertson}, B.~E., {Dalcanton}, J.~J., \& {de Blok},
  W.~J.~G. 2000, \aj, 119, 1579

\bibitem[{{van den Bosch} \& {Swaters}(2001)}]{Vand2001}
{van den Bosch}, F.~C. \& {Swaters}, R.~A. 2001, \mnras, 325, 1017

\end{thebibliography}
\end{document}